%% file: main.tex
\documentclass{neus2025}

\usepackage{graphicx}
\usepackage{soul}
\usepackage{url}
\usepackage[utf8]{inputenc}
\usepackage{microtype}
\usepackage{amsmath}
\usepackage{booktabs}
\usepackage{algorithm}
\usepackage[noend]{algorithmic}
\usepackage{amssymb}
\usepackage{mdframed}
\usepackage{xcolor}
\usepackage{pifont}
\usepackage{nicefrac}
\usepackage{makecell}
\usepackage{caption}

\usepackage{tikz}
\usepackage{import}
\usepackage{pgf}
\usepackage{wrapfig}

\usetikzlibrary{arrows,automata,positioning}

\title[Learning Formal Specifications from Membership and Preference Queries]{Learning Formal Specifications from Membership and Preference~Queries}
\usepackage{times}

\author{%
 \Name{Ameesh Shah} \Email{ameesh@berkeley.edu}\\
 \addr UC Berkeley
 \AND
 \Name{Marcell Vazquez-Chanlatte} \Email{marcell.chanlatte@nissan-usa.com}\\
 \addr Nissan Advanced Technology Center%
 \AND
 \Name{Sebastian Junges} \Email{sjunges@cs.ru.nl}\\
 \addr Radboud University%
 \AND
 \Name{Sanjit A. Seshia} \Email{sseshia@berkeley.edu}\\
 \addr UC Berkeley%
}

\begin{document}

\maketitle

\begin{abstract}
  Active learning is a well-studied approach to learning formal specifications, such as automata. 
  In this work, we extend active specification learning by proposing a novel framework that
  strategically requests a combination of membership labels
  \emph{and pair-wise preferences}, a popular alternative to membership labels. We formalize the notion of using preference queries in the context of specification learning by introducing \textit{Membership Respecting Preferences (MemRePs)}, a class of pair-wise preferences that can be used in conjunction with membership queries. 
  The combination of pair-wise preferences and membership labels allows for a more flexible approach to active specification learning, often reducing the number of membership queries required to learn specifications. We instantiate our framework for two different classes of specifications, demonstrating the generality of our approach.
  Our results suggest that learning from both modalities allows us to robustly and conveniently identify specifications via membership and preferences.
\end{abstract}
\begin{keywords}
  Preference-based learning, active learning, specification mining, automata learning
\end{keywords}

\input{commands}

\section{Introduction}
\label{introduction}
\input{introduction}



\section{Learning with Membership Respecting Preferences}
\input{problemstatement}

\section{Asking the right queries}
\label{sec:algorithm}
\input{alg}
\vspace{-1em}

\section{Experiments}
\label{sec:applications}
\input{applications}

\section{Related Work}
\input{relatedwork}

\section{Conclusion and Future Work}
\input{conclusion}

\acks{This work was supported in part by by the DARPA LOGiCS project under contract FA8750-20-C-0156, by Nissan and Toyota under the iCyPhy center, and DARPA contract FA8750-23-C-0080 (ANSR).}

\newpage
\bibliography{tacas22}

\newpage
\appendix
\input{appendix}

\end{document}

%% file: commands.tex
\newcommand{\Reals}{\mathbb{R}}

\newcommand{\MemRep}{\mathsf{MemRep}}
\newcommand{\universe}{\mathcal{U}}
\newcommand{\prefneq}{\prec}
\newcommand{\prefeq}{\preceq}
\newcommand{\po}{{\preceq}}
\newcommand{\teacher}{\mathcal{T}}
\newcommand{\mOracle}{\mathcal{M}}
\newcommand{\cOracle}{\mathcal{C}}
\newcommand{\concept}{\varphi}
\newcommand{\conceptClass}{\Phi}

\newcommand{\BB}{\{0, 1\}}
\newcommand{\states}{Q}
\newcommand{\alphabet}{\Sigma}
\newcommand{\transition}{\delta}
\newcommand{\initstate}{q_0}
\newcommand{\acceptstates}{F}

\newcommand{\yes}{\succ}
\newcommand{\no}{\prec}
\newcommand{\inc}{\|}

\newcommand{\eqdef}{\mathrel{\stackrel{\makebox[0pt]{\mbox{\normalfont\tiny def}}}{=}}}

\newcommand{\Distr}[1]{\mathrm{Distr}(#1)}

\newcommand{\sebastian}[1]{\textcolor{red}{SJ: #1}}
\newcommand{\savespace}[1]{\textcolor{red}{!}}
\newcommand{\ameesh}[1]{\textcolor{violet}{AS: #1}}
\newcommand{\marcell}[1]{\textcolor{blue}{MVC: #1}}

%% file: introduction.tex
An extensive body of work advocates for the use of \textit{formal logic} to model task specifications for learning-based autonomous agents~\citep{vazquezchanlatte2018learning, chou2020multistage, pan2023ltl, Baert2024learningtemporal}. These formal specifications provide a number of benefits relative to Markovian rewards: they offer a precise notion of satisfaction, are composable, and can be easily transferred and understood across environments. These benefits make logical specifications appealing in safety-critical applications such as planning, verification, and robotics~\citep{Webster20corroborative, yifru2023temporalconstraintpolicy}.

In practice, creating formal task specifications often requires intimate knowledge of both the environment and formal logic, which proves difficult for non-experts~\citep{hurley2024tricky}. To avoid this challenge, practitioners have developed methods to automatically learn formal specifications from data.
One popular approach is \textit{active} learning, where a non-human membership oracle labels generated agent trajectories as either positive (belongs in the set of desired behaviors) or negative~\citep{bastani2018active, bongard2005active, DBLP:journals/iandc/Angluin87}. We aim to extend these approaches to cases where we learn from non-expert teachers, such as humans.
While asking only membership queries suffices for completeness in active specification learning, we expect existing approaches to be challenging for human oracles, who are relatively bad at answering membership queries~\citep{DBLP:conf/rss/PalanSLS19,burton2021likert,Phelps2015Likert}. 

To address this challenge, we identify \emph{preference querying} as a promising alternative to membership querying, where the human is asked to rank two trajectories.  Preferences are relatively inexpensive to obtain, are generally preferred by human teachers, and are known to be less susceptible to mislabeling than membership query responses~\citep{DBLP:conf/rss/PalanSLS19,burton2021likert,Phelps2015Likert}. 
Moreso, the combination of the two signals is promising: \emph{Preference queries are comparatively more accurate but less informative than membership queries} ~\citep{DBLP:conf/rss/PalanSLS19}.

In this work, we contribute a general framework that allows for active learning from a combination of preferences and labeled examples.
The framework works as follows:
First, from existing known facts about the desired behavior, we generate candidate specifications that are consistent 
 with previously observed membership and preference constraints. 
Next, we use heuristics to generate a preference query and a membership query that will rule out as many incorrect candidate hypotheses as possible. We choose either to ask the preference query or membership query based on previous queries and the user's preferences for each query type. 
Then, we iteratively generate new candidate hypotheses that are consistent with the newly updated facts, and continue asking queries until the correct specification is found, if one exists.
To robustify our approach against a (limited) number of wrong answers, our algorithm can identify and ignore sets of answers that are inconsistent with one another (for example, preferring A to B and B to C but C to A).

In our experiments, we implement the aforementioned framework in two different classes of formal specifications, highlighting the generality of our approach. Our automated query selection process can avoid a substantial number of membership queries by asking additional informed preference queries. This trade-off between queries can be easily configured by setting the relative cost of answering each type of query to the teacher. 

\paragraph*{\textbf{Contributions}}
We present the formalism of membership-preserving preferences that allows for specification learning from oracles that can answer comparison and membership queries. 
To show the feasibility and efficacy of our formalism, we propose a \emph{concept class-agnostic} algorithm for querying oracles and run empirical evaluations on two different domains. Finally, we provide a novel method based on Boolean satisfiability (SAT) for identifying DFAs from labeled examples \emph{and} pair-wise preferences.

\color{black}


%% file: problemstatement.tex
We begin by developing the machinery to specify which behaviors within
a set are desirable, both in a relative and satisficing sense.
For a given \emph{universe}, $\universe$, containing \emph{atoms} $x \in \universe$, a formal specification, or \textit{concept} $\varphi \subseteq \universe$, contains a set of atoms. We denote by $\concept(x)$ the indicator, $[x \in \concept]$ and refer to a collection of concepts, $\conceptClass$, as a \emph{concept class}\footnotemark. W.l.o.g., we assume that the universe coincides with the union of the concepts in a concept class, $\universe = \bigcup \Phi$. 

\begin{example}
Concept classes and their universes can be finite or infinite. For
example, when representing formal task specifications,
one takes as the (infinite) universe, $\alphabet^*$, i.e., all words from a
finite alphabet $\alphabet$.
Similarly, languages represented by classes of automata, e.g., DFAs, are (infinite) concept classes.
\end{example}
\footnotetext{
For simplicity, we conflate a concept with its representation.
}

\begin{wrapfigure}[6]{r}{0.3\linewidth}
\vspace{-2em}
    \centering
     \input{hass_example}
     \vspace{-0.6em}
    \caption{A preference order over atoms, represented as a Hasse Diagram.}
\label{fig:hasse1}
\end{wrapfigure}

Next, we formalize the notion of pairwise preferences on atoms via
\emph{preorders}, $\prefeq$, on universes, i.e., transitive and
reflexive relations on $\universe$. Namely, we call a (fixed) preorder
on a universe a \emph{preference order} and interpret $x \prefeq y$ as
\emph{``$y$ may be preferred over $x$''}. We write $x \prefneq y$ if
$x \prefeq y$ and not $y \prefeq x$, interpreted as \emph{``$y$
  \emph{is} preferred over $x$''}. Two atoms have equal preference,
$x \equiv y$, if $(x \prefeq y) \wedge (y \prefeq x)$.  Finally, two atoms are \emph{incomparable}, $x \mid\mid y$, if
$\neg (x \prefeq y \vee y \prefeq x)$.

\begin{example}\label{ex:explicit_hass}
  An example preference order over the universe $\{a,b,c,d,e,f\}$, is
  shown in Fig~\ref{fig:hasse1}.  It is represented using a
  directed acyclic graph, $H = (V, E)$ called a \emph{Hasse diagram}.  The
  nodes of $H$ represent equivalence classes, i.e., for all $v \in V$,
  $x, y \in v \implies x \equiv y$, and the edges of $H$ represent
  strict preferences, i.e., $(x, y) \in E \implies x \prefneq y$.  The
  full preference order is the transitive reduction of $H$.
\end{example}
\begin{example}
  Costs and rewards offer a common way to define preference
  orders. For example, let $x \in \universe$ denote the set of paths
  through a maze and assign a cost, $c(x) \in \Reals$, to each path
  $x$ based on its length. A natural preference order is then given by
  comparing costs: $c(x) \leq c(y)$ implies $y \prefeq x$.
  This order is total (no atoms are incomparable) and illustrates that
  two distinct atoms can have equal preference, i.e.,
  $c(x) = c(y)$ implies $x \equiv y$, but not $x = y$.
\end{example}
\subsection{Learning with Preferences}
We now turn our attention to identifying an unknown concept,
$\concept^*$. Namely, we shall assume access to
a \emph{membership oracle},
$\mOracle \colon \universe \to \{\in, \notin\}$, to evaluate if
$x \in \varphi^*$, as well as a comparison oracle,
$\cOracle \colon \universe^ 2 \to \{\prec, \|,
\succ, \equiv\}$, to provide preferences between atoms,
e.g., $[C(x, y) =\hspace{0.2em} \prec]$ iff $[x \prec y]$.
Invocations of these oracles are \emph{queries}.

\begin{mdframed}[backgroundcolor=black!10,nobreak]
  \textbf{Problem Statement:} Let $\varphi$ be an \emph{unknown}
  specification in concept class $\Phi_\text{init}$.  Given membership
  and comparison oracles $\mOracle$ and $\cOracle$, infer $\varphi$.
\end{mdframed}
\begin{remark}
For finite concept
classes it suffices to consider only the membership oracle $\mOracle$, ignoring query and time complexity. In particular,
one may pair-wise consider all concepts and pose a membership query from the symmetric difference of the concepts to uniquely identify the concept.
However, for many domains~\citep{burton2021likert}, obtaining accurate membership oracles is expensive.  The key question in this work is
\emph{how to exploit the availability of the comparison oracle} $\cOracle$.
\end{remark}
\input{memrep}
%

%% file: hass_example.tex
\begin{tikzpicture}
  \node[draw,circle,fill,inner sep=1pt,label={east:{\scriptsize f}}] at (2,0) (f) {};
  \node[draw,circle,fill,inner sep=1pt,label={south:{\scriptsize e}}] at (1.35,-0.5) (e) {};
  \node[draw,circle,fill,inner sep=1pt,label={south:{\scriptsize d}}] at (0.65,-0.5) (d) {};
  \node[draw,circle,fill,inner sep=1pt,label={north:{\scriptsize c}}] at (1.35,0.5) (c) {};
  \node[draw,circle,fill,inner sep=1pt,label={north:{\scriptsize b}}] at (0.65,0.5) (b) {};
  
  \node[draw,circle,fill,inner sep=1pt,label={west:{\scriptsize a}}] at (0,0) (a) {};
  
  \draw[->] (a) -- (b);
  \draw[->] (b) -- (c);
  \draw[->] (c) -- (f);
  \draw[->] (a) -- (d);
  \draw[->] (d) -- (e);
  \draw[->] (e) -- (f);
  
  \draw[dashed,thick,color=blue!60] (2,-0.3) -- (0,-0.3);
  \node[color=blue!60] at (2,-0.6) {\footnotesize $\concept_2$};
  
  \draw[dashed,thick,color=red!60] (1.3,0.9) -- (0,-0.7);
  \node[color=red!60] at (1,0.9) {\footnotesize $\concept_1$};
  
\end{tikzpicture}


%% file: memrep.tex
To leverage the comparison oracle,
we  relate preferences to membership in $\concept^*$.  We therefore focus on preference orders that respect membership: atoms outside of the concept, $x \not\in \concept^*$, cannot be preferred to
atoms in the concept, $y \in \concept^*$.

\begin{definition}
A preference order is a \emph{membership-respecting preference} (MemReP) w.r.t.\ $\concept$ if
\begin{equation}\label{eq:weakMemReP}
x \prefeq y  \implies  \concept(x) \leq \concept(y).
\end{equation}

\end{definition}

\begin{example}\label{ex:extreme}
  All preference orders are membership respecting w.r.t. concepts $\top \eqdef \universe$ and $\bot \eqdef \emptyset$.
  Similarly, any oracle that yields $C(x, y) = \inc$ for all $x,y \in \universe$ is
  vacuously membership-respecting.
\end{example}
\begin{example}\label{ex:total}
  Cost-based preferences, like in our path example, together with
  thresholded cost concepts, i.e.,
  $
     \concept_\delta \eqdef \{x~\in~\universe\ :\ c(x) \leq \delta\},
  $
  for some cost map $c \colon \universe \to \Reals$ and threshold $\delta \in \Reals$, 
  are membership respecting.
\end{example}
\begin{example}\label{ex:pref_id}
  We continue with Ex.~\ref{ex:explicit_hass}. This order is
  membership-preserving for $\concept_1 = \{ a, b \}$, but not for
  $\concept_2 = \{ d, e \}$. Graphically, $\concept_2$ is not
  membership-preserving as there is an edge from $\bar{\concept_2}$ to
  $\concept_2$.
\end{example}

Example~\ref{ex:pref_id} illustrates that the MemReP
assumption is sometimes strong enough to distinguish concepts without
any membership queries. Namely, if there exists a pair of atoms,
$x \in \concept_1 \setminus \concept_2$ and
$y \in \concept_2 \setminus \concept_1$, such that $\cOracle(x, y) \neq ||$, then $\cOracle(x, y)$ is 
guaranteed to distinguish $\concept_1$ and $\concept_2$ under the
MemReP assumption.

\begin{definition}
Let $X = (X_\mOracle, X_\cOracle)$ be a tuple of sets of membership and comparison labeled examples respectively, e.g., 
\begin{align*}
X_\mOracle = \{(x_1, \in),\ldots, (x_j, \notin)\} \\
X_\cOracle = \{(x_k, y_k, \|), \ldots, (x_n, y_n, \prec)\} .
\end{align*}
A concept $\concept$ is \emph{consistent} with $X$ if (i) $\concept$ agrees with all membership assignments in $X_\mOracle$, e.g., $(x, \in) \in X_\mOracle \implies x \in \concept$; and (ii) All preferences in $X_\cOracle$ are membership respecting under $\concept$, e.g., $(x, y, \prec \concept(x) \leq \concept(y)$. Given a concept class $\Phi$,
we denote by $\Phi^X$ the set of all concepts in $\Phi$ consistent with $X$.\end{definition}
\subsection{Abstract Algorithm}

We now return to the problem of identifying an unknown concept given
access to a membership oracle and a membership respecting comparison
oracle.  In Alg.~\ref{alg:generic_alg}, we outline the general process for learning concepts from
\emph{finite} concept classes given such oracles. The learner begins with a concept class, asks a membership or comparison query, and then removes concepts that are inconsistent with the query result until a unique concept remains.
\vspace{-0.5em}
\begin{algorithm}[H]
  \caption{Generic algorithm for concept identification.\label{alg:generic_alg}}
  \begin{algorithmic}[1]
  \STATE Assign $\Phi$ as the initial concept class $\Phi_\text{init}$.
  \STATE Initialize the set of labeled examples $X$ as $(\emptyset, \emptyset)$.
  \WHILE {$|\Phi| > 1$}
    \STATE Ask either a membership or a comparison query given $X$ and $\Phi^X$.
    \STATE Add the result to $X$.
  \ENDWHILE
  \OUTPUT $\Phi^X$\hfill\COMMENT{$\Phi^X = \emptyset$ or $\Phi^X = \{\phi^*\}$.}
  \end{algorithmic}
\end{algorithm}

\begin{proposition}\label{prop:terminate}
  Suppose for every iteration, the probability of asking a distinguishing membership query,
  i.e., asking $\mOracle(x)$ for $x$ in the symmetric difference of two concepts in $\Phi^X$, is bounded from below. Then, Alg~\ref{alg:generic_alg} almost surely terminates.
\end{proposition}
\begin{remark}
  To treat infinite concept classes, we appeal to Occam's Razor, and
  seek to find the ``simplest" concept that is consistent with the
  data. Formally, we assume that the finite concept class is the
  (countably infinite) union of finite concept classes,
$
    \Phi = \bigcup_{i=1}^{\infty} \Phi_i,
$ where $i$, is taken as a complexity measure, e.g, number of states of
an automaton. We extend the above process by seeking to find the
smallest $i$ such that the above process returns a singleton. Finally,
as is standard in this setting, we will additionally assume access to
an equivalence oracle to provide completeness guarantees. In practice,
such equivalence queries are often impractical, and are approximated
via conformance testing and sampling.
\end{remark}
In order to realize this algorithm in practice, we require three ingredients. First, we must develop methods that can synthesize a consistent concept over both membership query and preference query results for a specific concept class. This operation enables symbolically interacting with $\Phi^X$ in Alg.~\ref{alg:generic_alg}. We provide synthesis methods for the concept classes used in our experiments in the appendix. Second is an intelligent means to select a query for $\mOracle$ or $\cOracle$, which we expand on in the sequel. Last is a means to non-trivial possibility of labeling mistakes, which we provide in the appendix. 


%% file: alg.tex
We now discuss a concept class agnostic strategy to select queries for an efficient version of Alg.~\ref{alg:generic_alg}.
\subsection{Cost model}
Before we optimize our algorithm, we must set the measure that we aim to optimize.
In any active learning algorithm, the selection of the queries is central to its performance.
In particular, we seek to balance minimizing the number of queries asked and computational costs.
To this end, we model the costs of preference and membership queries to be time invariant and constant, based on a weighted sum:
\begin{equation} \label{eq:costfunction}
    \text{cost}(\text{queries}) \triangleq a \cdot \# \text{mem} + b \cdot \# \text{pref},
\end{equation} where $\#$mem and $\#$pref refer to the number of membership and preference queries and $a, b \in \mathbb{R}_\infty$. 
For example, $a=b$ treats membership and comparisons interchangeably 
and $a=\infty, b=1$ lexicographically prefers comparisons over membership queries\footnote{Our algorithm can handle arbitrary cost models over $\#$mem and $\#$pref and is not dependent on the specific structure of equation \ref{eq:costfunction}.}.
\subsection{Contextual Bandit Formulation}
Because the results of the queries are a priori
unknown, na\"ively optimizing a given cost model is often infeasible.
Furthermore, as the next example illustrates, adversarial teachers can
induce arbitrary regret.
\begin{example}\label{ex:adversarial_teacher}
  Recall the lexicographic example, $a=\infty, b=1$. If all
  preferences yield incomparable, then one would regret making any
  comparison queries. On the other hand, if one ignores comparisons,
  but the underlying preference lattice is total as in Ex.~\ref{ex:total}, then one may ask
  many more membership queries than is  required. In
  particular, if the (total) preference order were known, then the learner
  could binary search (using $\mOracle$) for the unique point
  where membership changes.
\end{example}
Further, we highlight that, even if a model for query responses is
known, optimally planning a sequence of queries is often
computationally intractable.  There are
$M = \binom{|\universe|}{2} + \universe$ queries to ask in a single
step, and thus roughly $3^{M^t}$ possible combinations when asking up
to $t$ queries. Thus, even assuming $\universe$ is finite, but
non-trivially small, e.g., strings of length at most 10, the search
space quickly becomes intractable.

Thus, we propose a two stage formulation to optimize the query selection based on adversarial
contextual multi-armed bandits (CMABs)~\citep{DBLP:journals/siamcomp/AuerCFS02}, specifically, multi-armed bandits with expert advice. In this formulation, the classic multi-armed bandit framework is extended by introducing \textit{experts} that consider the arms and context in the problem, and provide recommendations in the form of probability distributions over the arms that our algorithm can then make use of in its policy. In order to instantiate our CMAB algorithm with experts, we make the following choices:
(i) We use a heuristic to select candidate comparison and
membership queries, the \emph{arms} in our formulation. 
The heuristic is a (Monte Carlo) estimate of the concept class size's reduction for each query's possible outcome. (ii) We define a (bounded) proxy cost per arm weighing
the query cost against the (estimated) reduction in the concept class:
\begin{equation}\label{eq:loss}
\text{loss}_c \eqdef \frac{c}{\max(a, b)} \cdot \frac{|\Phi'|}{|\Phi|} \in [0, 1],
\end{equation}
where $c$ is the cost of the selected arm's query type (thus either $a$ or $b$).
Finally, (iii) we encode two heuristic query strategies as \emph{experts}
assigning probabilities to each arm, described in Section \ref{sec:experts}.  
The resulting CMAB game proceeds in rounds described in Alg.~\ref{alg:CMAB}, each round corresponds to an iteration in  Alg.~\ref{alg:generic_alg}.
\begin{algorithm}[t]
    \caption{CMAB round for selecting a query.}
    \label{alg:CMAB}
   \begin{algorithmic}[1]
\STATE Select comparison and membership queries via Alg.~\ref{alg:query_selector}.
\STATE Estimate the loss~\eqref{eq:loss} for different query outcomes.
\STATE Provide average and worst-case expert advice distributions on the arms (queries), $E_1, E_2$, see Sec.~\ref{sec:experts}. 
\STATE Sample which expert, $E \sim \{E_1, E_2\}$, to listen to based on the historical performance of the expert.
\STATE Sample the arm (query) based on $E$'s arm distribution.
\STATE Compute the actual loss and update expert distributions.
   \end{algorithmic} 
\end{algorithm}

\begin{algorithm}[t]
\caption{Query selection heuristic.\label{alg:query_selector}}
\begin{algorithmic}[1]
  \STATE Select a set of up to $\alpha$ unrefuted concepts, 
    $\Psi \subseteq \Phi$.
  \STATE Let $X$ be a set of atoms such that (i) $|X|\ \in [\alpha, \beta]$ and (ii) $X$ distinguishes concepts in $\Psi$, i.e. \[\forall \concept_1, \concept_2 \in \Psi~.~\exists x \in X~.~ x\in \concept_1 \Delta \concept_2.\]
  \vspace{-1.5em}
  \STATE Find $x \in X$ that minimizes
  the (worst-case) number of concepts in $\Psi$ that are consistent after $\mOracle(x)$.
  \STATE Find $y, z \in X$ that minimize
  the (worst-case) number of concepts in $\Psi$ that are consistent after $\cOracle(y, z)$.
  \STATE Return candidate queries $\langle\mOracle(x), \cOracle(y, z)\rangle$.
\end{algorithmic}
\end{algorithm}

To avoid considering all atoms in $\universe$, we propose the
 concept class agnostic heuristic in Alg.~\ref{alg:query_selector} to select candidate
membership and comparison queries. For our implementation, we use $\alpha = 2$, and take at most two atoms per concept, i.e., $\beta = 2\alpha$.


\vspace{-1em}
\subsection{Worst and average case advice}\label{sec:experts}
We propose to give advice towards either of the two queries based on combining two perspectives.
In particular, we propose to use the following two experts with the following advice arm distributions:  \textbf{(1) Pessimistic}: Weighs each arm by its worst-case loss. Ignores the incomparable answer for preference queries. \textbf{(2) Historical}: Weighs each arm by its expected loss, computed by averaging the previous losses incurred by pulling that particular arm.
    
To construct the advice distribution, we take the softmax of the weights of the comparison and membership queries.
  The expert selection distribution is then updated using the standard exp4 contextual
  bandit algorithm~\citep{DBLP:journals/siamcomp/AuerCFS02} which
  guarantees that we will switch between our pessimistic and
  historical arm strategy in a manner such that, with respect to loss~\eqref{eq:loss}, the algorithm would not have done much better sticking to advice from a single expert.
  
Let us share the following intuitions for our choice of experts. 
The pessimistic expert helps to ensure some notion of progress. However, as an incomparable answer for the comparisons never yields any progress, we exclude this case from the expert. 
On the other hand, we have seen in Ex.~\ref{ex:adversarial_teacher} that the utility of a comparison
query depends heavily on the oracle's underlying preference order. To capture this observation, 
the historical expert learns whenever almost all comparisons yield incomparable and will promote using membership in these cases. Likewise, this expert will discover when there is a total preference order and all atoms are comparable, and perhaps even equivalent.  
Finally, observe the following proposition about the proposed algorithm:
\begin{proposition}\label{prop:completeness}
Assume the unknown concept, $\varphi^*$, is in $\Phi$. Running Alg~\ref{alg:generic_alg} using Algs~\ref{alg:CMAB},\ref{alg:query_selector} for queries (almost surely) identifies $\varphi^*$ after finitely many queries.
\end{proposition}
A proof sketch for Prop.~\ref{prop:completeness} is provided in the appendix.


%% file: applications.tex
In this section, we instantiate Alg.~\ref{alg:generic_alg} on two families of concept classes, demonstrating the flexibility and efficacy of the proposed formalism and heuristics. 

\vspace{-1em}
\subsection{Deterministic Finite Automata}
\label{sec:encoding}

\input{identification}
\vspace{-2.0em}
\subsection{Monotone Predicate Families}
\label{sec:thresholds}
\input{thresholds}

%% file: identification.tex
To begin, we apply the algorithm and operations outlined in
Sec.~\ref{sec:algorithm} to the space of DFA identification, 
relying on the performance of modern Boolean satisfiability (SAT)
solvers.
\begin{wrapfigure}[7]{r}{0.26\linewidth}
\vspace{-1em}
  \centering
\scalebox{0.8}{
\begin{tikzpicture}
  \node[circle,draw] (s1) {};
  \node[circle, initial,draw, above=of s1, initial where=left, initial text=] (s2) {};
  \node[circle,draw=green!50!black, accepting, right=of s2] (s3) {};
  \node[circle,draw,right=of s1] (s4) {};
  
  \draw[->] (s1) edge[bend left] node[left] {\textcolor{blue}{Bl}} (s2);
  \draw[->] (s2) edge[bend left] node[left] {\textcolor{brown}{Br}} (s1);
  \draw[->] (s1) -- node[auto,sloped] {\textcolor{red}{R}\textcolor{yellow!60!black}{Y}} (s4);
  \draw[->] (s2) -- node[auto,sloped] {\textcolor{yellow!60!black}{Y}} (s3);
  \draw[->] (s2) -- node[auto] {\textcolor{red}{R}} (s4);
  \draw[->] (s3) -- node[auto] {\textcolor{red}{R}} (s4);
\end{tikzpicture}
}
\vspace{-0.5em}
  \caption{A DFA encoding the considered task.}
  \label{fig:target-dfa}
\end{wrapfigure}
A \emph{Deterministic Finite Automaton}, DFA, is a 5-tuple, $(Q, \Sigma, \delta, q_0, F)$, consisting
of a finite set $Q$ of \emph{states}, a finite \emph{alphabet} $\Sigma$, a \emph{transition function}, $\delta \colon Q \times \Sigma \to Q$, an \emph{initial state} $q_0$, and the \emph{accepting states} $F \subseteq Q$.
The function $\delta^* \colon \Sigma^* \to Q$ denotes the lifting of $\delta$ to sequences of symbols (strings), via repeated application.
Finally, the language of DFA $\mathcal{D} = \{ w \in \Sigma^* \mid \delta^*(w) \in F \}$ is set of strings that reach an accepting state, $q\in F$.

\paragraph{Learning Task Specifications for Robots}
We study the problem of conveying a task to an agent (robot)
moving about an environment. For example, consider the task specification
that says ``avoid red (lava) tiles and do reach a yellow (recharge) tile (RY)'' and ``between visiting blue (water) tiles and yellow tiles, the agent must be visit
a brown (dryer) tile. (BBY)'' from \cite{vazquezchanlatte2018learning}.
This task specification is described by the DFA shown in Fig~\ref{fig:target-dfa}.
We assume that the robot is pre-programmed to universally assume that a task will
contain the RY constraint and seeks to interactively learn the domain-specific BBY constraint (in the form of a DFA) from a user.

First, observe that because DFAs are closed under conjunction, a new DFA can be easily derived that will not violate the a priori specified RY task.
Further, depending on the system
dynamics, the user may prefer some traces to others, and is able to provide these pair-wise comparisons. Thus, we pose this task learning process as an active learning problem, where the user can provide (i) \emph{pair wise
preferences} and (ii) \emph{membership query responses (labels)}
that specify whether an example is good or bad. We want to leverage that query mechanism to potentially
accelerate or robustify the learning of the DFA. 

The oracle uses a
randomly generated membership respecting preference order, where
approximately $\nicefrac{1}{10}$ of the comparisons yield
incomparable.  To refute equivalence queries, we sample a random
string from the symmetric difference of the true DFA and the current
hypothesis. Finally, we fix the comparison cost to 1 and vary the membership
cost to study the trade off in queries the algorithm makes.

\begin{figure}[t]
  \centering
  \begin{minipage}{0.48\textwidth}
  
  \includegraphics[width=0.98\textwidth,height=0.6\textheight,keepaspectratio]{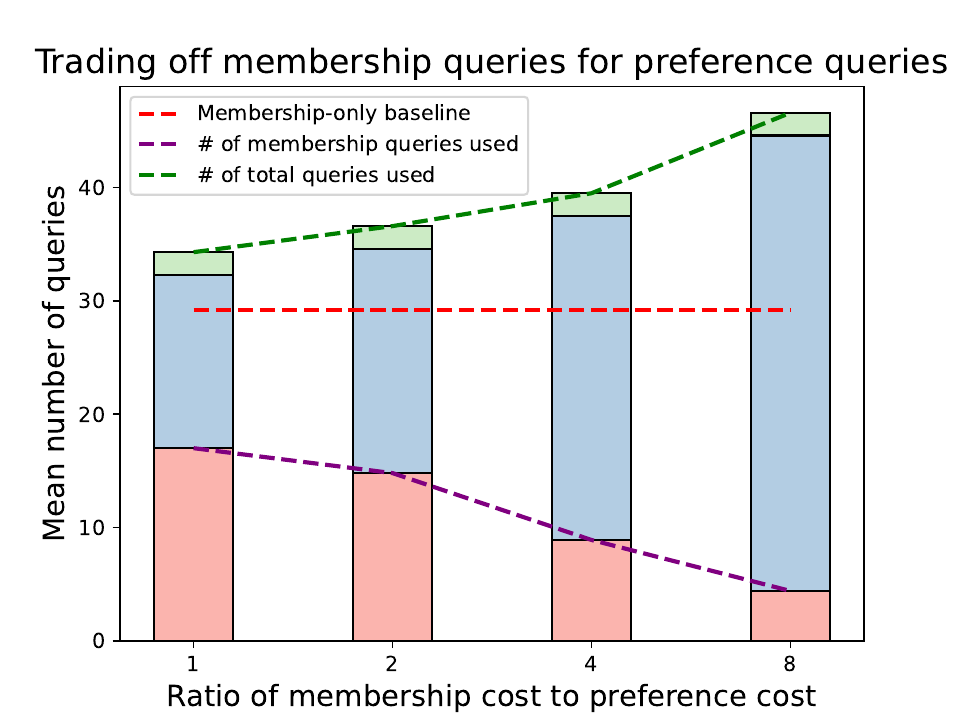}
  
  \end{minipage}
    \begin{minipage}{0.48\textwidth}
  
  \includegraphics[width=0.98\textwidth,height=0.6\textheight,keepaspectratio]{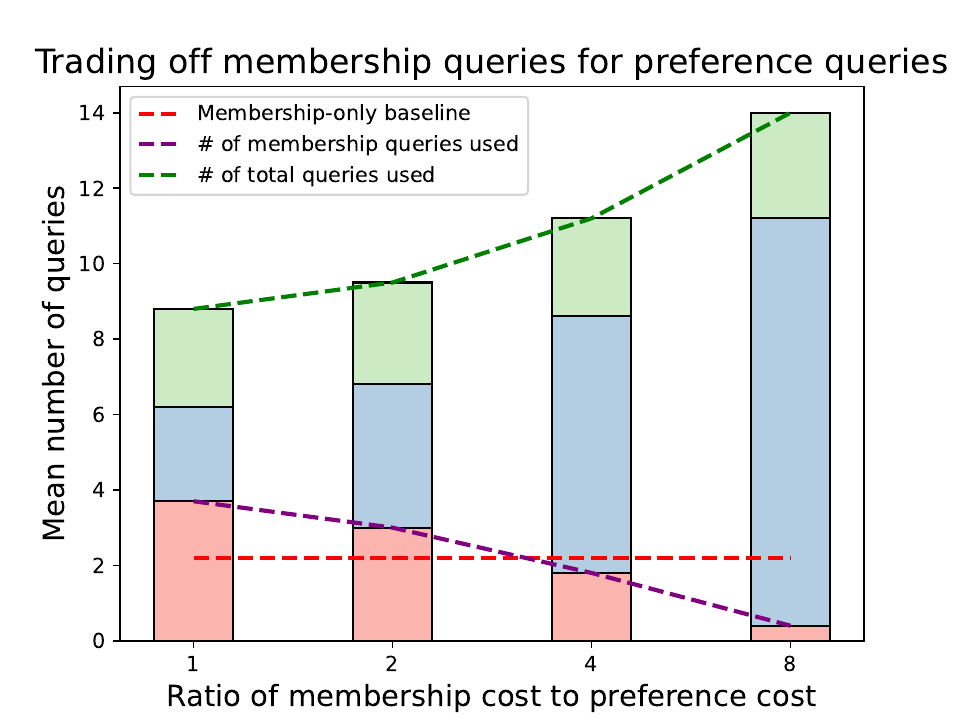}
  \end{minipage}

  \vspace{-0.5em}
\caption{Trade-off between preference
queries and membership queries in the DFA
domain (Left) and the Monotone Predicates domain (Right). The bars plotted show the contribution of
membership (red, bottom), preference (blue,
middle), and equivalence (green, top) queries.} 
  \label{fig:pref_mem}
  \vspace{-1em}
\end{figure}

The results of the experiment are shown in Fig~\ref{fig:pref_mem}. We
observe that, as expected, as the membership cost increases (i) the
total number of queries increases (ii) the total number of membership
queries decreases. Specifically, \emph{removing 1 membership query adds an average of 2.03 preference queries} each time the cost doubles. Additionally, the use of preference queries allows the use of membership queries to be well below
the baseline rate of setting the costs such that only membership
queries are used; that is, setting the cost of a preference to $\infty$.

Finally, we compare with the discriminate tree variant of
L*~\citep{DBLP:books/daglib/0041035,DBLP:journals/iandc/Angluin87}, a
classic algorithm for learning DFAs from membership and equivalence
queries. We use the AALpy  library~\citep{aalpy} implementation of L*, and find that the implementation requires 11 membership queries and 2
equivalence queries (the same number as Fig~\ref{fig:pref_mem}) that
cannot be answered by the RY task prior knowledge. While our algorithm requires more overall queries, we observe that the number of membership queries can be
fewer than 11 as at membership cost 4, all without concept class specific
tailoring. Furthermore, we remark that these two algorithms could be used in concert by adapting the membership query selection to use L* provided
queries.

\paragraph{Additional Experiments} 
In order to further study the performance of our algorithm in learning DFAs, we performed the following studies: First, we applied our algorithm to learning the Tomita Language DFAs, a standard benchmark in DFA learning, to evaluate performance across a diversity of DFAs. Next, we evaluated our algorithm on classes of DFAs that vary in size to evaluate how our algorithm scales as the target DFA increases in complexity. Finally, we implemented and evaluated a method in our algorithm that can identify and tolerate incorrect responses from our oracle. 
The full details and results of these experiments are available in the appendix.

\


%% file: thresholds.tex
Next, we study monotone predicate families. A \emph{monotone predicate family} is a concept class with an (arbitrary but fixed) partial order $\sqsubset$ defined over the concepts such that $\concept \sqsubset \concept'$ implies  $\concept \subseteq \concept'$. 
Increasing a concept thus monotonically increases the set of atoms included by the concept.
We motivate studying monotone predicates using a series of motivating examples.
\begin{example}
\label{ex:car}
We consider the on-boarding process of a hypothetical car that queries the user to learn what (safe) distances to other objects they deem comfortable. 
In the scope of this paper:
(a) The resulting behavior should \emph{never} violate any pre-defined
  safety constraints.
(b) The on-boarding experience should be brief, i.e., the system should try to minimize the number of queries.
(c) Communication should be unambiguous and concrete to cover edge cases.
\end{example}
The above example can be cast as a 2-dimensional monotone
predicate family, where one dimension corresponds to maximum time,
$\tau \in [0, T]$, the user is willing to wait to reach the
destination and the other dimension corresponds to the minimum
distance, $d \in [0, D]$, to another car the user is comfortable
with. The corresponding partial order has $(\tau, d) \sqsubset (\tau', d')$ if $\tau < \tau'$ and
$d > d'$.

We seek to understand the trade-off between membership
and comparisons incurred by our algorithm given different membership
and comparison costs.  
We instantiated our algorithm with a size-indexed concept class,
where each $\Phi_i$ corresponds to a uniform 2d grid of parameters with $i$
points per axis. Equivalence queries provide a labeled bi-partition
separating the concept.
Like with our DFA experiment, the preference order was randomly generated such that approximately $\nicefrac{1}{10}$ on atoms are incomparable and approximately $\nicefrac{1}{3}$ of atoms whose preference is not forced by the MemReP condition are strictly ordered.
Furthermore, as a baseline, we compare against an learner that only uses
equivalence and membership queries by setting the comparison query cost
sufficiently high. 

Fig.~\ref{fig:pref_mem} shows the averaged results on 100 randomly generated concepts with the comparison cost fixed to 1 and the
membership cost changing on the horizontal-axis. The vertical axis corresponds to the average of number of queries across all preference orders.
First, the average number of equivalence queries was the same for each instance on the same concept, including the shown membership only base-line. 
Furthermore, as desired, increasing the cost of the membership queries results in the average
number of membership queries decreasing, at the expense of additional
comparison queries: as the relative cost doubles, \textit{removing 1 membership query adds on average 2.4 membership queries}.
This is expected given that comparisons provide less information
about the concept's label than a membership query. 

We also observe that initially, introducing preferences occasionally \emph{increases} the number of membership queries. This effect is due to: (i) the greedy nature of our algorithm, meaning that ``good'' membership queries are ignored as they correlate with ``good" preference queries, and (ii) our hyperparameters - namely the temperature of the softmax in the expert advice - was tuned with the assumption that membership would cost significantly more than preferences.



%% file: relatedwork.tex


\paragraph{Active learning of rewards using preferences.} Inverse reinforcement learning (IRL)~\citep{ng2000Algorithms,abbeel2004apprenticeship} often relies on high-quality demonstrations to learn a reward function.  More recently, works have  proposed approaching IRL from an active learning perspective, asking a teacher for information-rich feedback on learner-generated examples, such as corrected examples or labels on sections of generated examples~\citep{hadfield2016cirl,brown2018activeIRL}. Preference-based reward learning has emerged from these active approaches as a popular method due to the accuracy and relative inexpensiveness of preference queries~\citep{Holladay-2016-5540,DBLP:conf/nips/WilsonFT12}.
To overcome the limited information content gained from relative comparisons, various techniques  have been devised to actively select preference queries that maximize the amount of information gained~\citep{Sadigh2017ActivePL,biyik2018batch,xu2017noisetolerant,basu2019hierarchicalqueries, xu2020thresholding}, typically by removing maximal volume from the hypothesis space.

\paragraph{Grammatical inference and concept learning.}
Grammatical
inference~\citep{de2010grammatical} refers to the rich
literature on learning a formal grammar (often an automaton) from data.
Examples include learning the smallest automata consistent
with a set of positive and negative
strings~\citep{de2010grammatical}, learning an
automaton using membership and equivalence
queries~\citep{DBLP:journals/iandc/Angluin87}, and extensions that relax the assumption of equivalence oracles and the ability to ask any membership query~\citep{moeller2023incomplete}. 
Within this literature, the key contributions of
our work are to take into account preferences via the concept-agnostic framework. Most similar to our work is~\citet{Hsiung2023rmpref}, which uses preferences within the L* framework to learn reward machines, a specialized type of automaton. This work focuses on preference-based specification learning in a more specific setting and is complementary to our contribution.
Finally, our algorithm can be seen as an extension of version space learning~\citep{DBLP:conf/ictai/SverdlikR92}, where we use preference-based learning to build on existing methods that leverage either explicit data structures or SAT-based DFA-identification~\citep{DBLP:conf/lata/UlyantsevZS15,HeuleV10} to realize candidate elimination. 


%% file: conclusion.tex
In this paper, we present a generic framework for learning task specifications (concepts) from actively acquired (noisy) preferences and labeled examples.
Despite being concept class agnostic, we demonstrated the efficacy of our approach on two very different concept classes.

Nevertheless, interesting future work includes considering principled approaches to deriving domain-specific optimizations of the heuristics used in our framework. 
Furthermore, we hope to consider  more expressive concept class families such as symbolic automata and context-free grammars: With symbolic automata we hope to support a mix of thresholded rewards and DFAs. 

%% file: appendix.tex
\clearpage
\section*{Appendix}

\renewcommand{\thesubsection}{\Alph{subsection}}

\subsection{MemRePs for DFAs}

In order to realize our algorithm for DFA learning, we need to support
synthesizing a DFA consistent with \emph{labeled examples} $V_+, V_-$ and an observed set of preferences, $V_\prec$, i.e. the results of previously observed membership and comparison queries respectively.

We extend the SAT encoding presented in
\cite{DBLP:conf/lata/UlyantsevZS15,HeuleV10} for the passive
identification of DFAs from positive and negative examples to support
membership respecting preferences.  The encodings operate by using the
provided positive and negative examples to form a prefix tree. The
nodes indexed by positive and negative examples are annotated with
whether they accept or reject. Two states can be merged if
they are indistinguishable in the resulting transition system. This
feature, together with the determinism of a DFA, are captured by
transforming the problem into a k-color graph coloring problem, where
$k$ is the fixed size of the DFA that is to be identified. The resulting
graph coloring problem is then encoded as a Boolean satisfiability (SAT) query.
More specifically, for each labeled word, $v$, the
SAT encoding includes (i) a variable
$x_{v, i}$ indicating if word $v$ accesses state (color) $i$ and (ii)
a variable $z_i$ indicating whether state (color) $i$ is accepting.

\subsubsection*{Encoding}

We extend the existing encoding to incorporate membership respecting
preferences. 
Using these variables, a membership preserving preference,
$(w, v) \in V_\prec$, can be encoded using the following constraints:
\begin{equation} \label{preference_constraint}
 \forall i,j~.~(w \prec_\varphi v) \in V_{\prec}:\quad \hspace{-1.5em}
 \underbrace{(x_{w,j} \wedge x_{v,i} )}_{v~\&~w \text{ access }i~\&~ j} \hspace{-0.5em} \implies \hspace{-1.5em}\underbrace{(z_j \implies z_i)}_{z_j \implies z_i \text{ is equiv to } z_j \leq z_i}   
\end{equation}
This constraint formalizes the previously mentioned logic that non-preferred trajectories' acceptance will lead to preferred trajectories' acceptance, and preferred trajectories' rejection will lead to non-preferred trajectories' rejection. If this were not the case, the MemReP condition would lead to a contradiction.

\subsection{MemRePs for Monotone Predicate Families}




Recall the motivating example mentioned in section~\ref{ex:car} that we cast as a 2-dimensional monotone
predicate family. We provide additional insight regarding our algorithm in the context of monotone predicate families as follows.
\paragraph{Geometric Perspective}
In order to study the generic behavior of our algorithm on monotone
predicate families, it helps to focus on a geometric
interpretation of the concept class. For that, we assume that we adequately parameterize the concepts in a concept class $\Phi = \{ \varphi_\theta \mid \theta \in [0,1]^d\}$. The parameters $\theta$ induce a natural pointwise (or product) order $<$ with $\theta < \theta'$ if for all $i < d~.~\theta_i < \theta'_i$. It is then straightforward to define the order on the concepts as $\varphi_\theta \sqsubset \varphi_{\theta'}$ if $\theta < \theta'$.
Next, observe that any atom, $x \in \universe$, partitions the parameter space into two
regions, 
\begin{equation}
  \Theta^+_x \eqdef \{\theta \mid x \in \concept_\theta\}\hspace{3em}
  \Theta^-_x \eqdef {[0, 1]}^d \setminus \Theta^+,
\end{equation}
called a \emph{monotone bi-partition}.
With this perspective, membership is entirely determined by whether or not the underlying
concept's parameter, $\theta$, is
in the accepting set,
\begin{equation}\label{eq:proxy_accept}
  x \in \concept_{\theta} \iff \theta  \in \Theta^+.
\end{equation}

\begin{figure}[t]
  \centering
    \centering \scalebox{0.7}{
    
    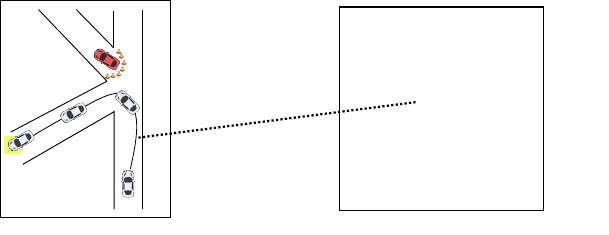 }

  \caption{Mapping Ex.~\ref{ex:car} to geometric perspective of concept class. 
  \label{fig:geometric}}
\end{figure}

\begin{example}
For instance, scaling and reflecting the parameters,
  $\theta = \left(\frac{\tau}{T}, \frac{D - d}{D}\right)$,
  yields a monotone parametric family. Fig~\ref{fig:geometric} illustrates.
  In particular, the ego (white) car wants to go to the yellow region within some time budget, while maintaining a minimum distance to the red parked car. This gets mapped to a bi-partition in the normalized parameter space.
\end{example}

\paragraph{Thresholded rewards}
Monotone predicate families include task specifications as thresholded rewards to weighted sums over multi-dimensional rewards. In particular, let $\vec{f}(i) \in {[0, 1]}^{d }$ be a sequence feature vector for
  $i \in [1, N]$, for $N \in \mathbb{N}$. 
Thresholded sums of linear rewards on these feature vectors can be cast as a $d+1$ dimensional monotone
  predicate family. In particular, let the thresholded sums be defined as $\sum_{i=1}^N w \cdot f(i) > \delta$ for $w \in {[-1, 1]}^{d}, \delta \in [-d, d]$. We may define $\theta$ using:
  \begin{equation}
    \theta_j =
    \begin{cases}
      (1 - w_j)/2  & \text{if } j \leq d\\
      (\nicefrac{\delta}{d} + 1)/2 & \text{if } j = d + 1
    \end{cases}.
  \end{equation}
 \begin{remark}
  If the $\cOracle(x, y)$ is total, one derives a (noiseless) variant of the learning setting considered in~\cite{xu2020thresholding}.
 \end{remark}

\subsection{Proof sketch for Proposition~\ref{prop:terminate}}
\begin{proof}[Sketch]
   Under exp4, a series of unproductive preference queries, i.e., ones that do not change $\Phi^X$, will exponentially increase the weight of the historical expert. Similarly, the historical expert will exponentially increase the weight of the distinguishing membership query arm. Finally, because the per round loss is bounded, there exist a lower bound on asking  distinguishing membership query. By Prop~\ref{prop:terminate} the algorithm almost surely requires finite queries.
\end{proof}

\subsection{Handling Error}
\label{sec:errors}
\input{errors}

\subsection{Learning DFAs: Additional Details and Results}\label{sec:appendix_dfa}
\subsubsection{Tomita Languages}\label{sec:tomita} The Tomita languages \cite{Tomita1982LearningOC} are a standard set of regular languages and DFAs frequently used as a benchmark in DFA learning and identification. The seven languages have a number of appealing qualities: they are relatively parsimonious, and they collectively span a number of interesting properties, including distributions of accepting and rejecting strings, existence of sink states, and relative ease of identification with a small number of membership queries.

For the Tomita languages, we used a manually designed preference ordering to incoporate more semantic meaning into the ordering and thereby encourage better learning from preference queries themselves. The preference ordering works as follows: Positively labeled atoms are still always preferred over negatively labeled atoms, as expected. When comparing two negatively labeled atom, the atom that took longer to reach a sink state (i.e., a rejecting state that cannot be transitioned out from) in the DFA was preferred, if a sink state existed. If neither atom reached a sink state, the atom with a longer accepting prefix was preferred. When comparing two positive atoms, we enumerate the four possible two-token extensions to the atoms and prefer the atom that is more frequently accepted when considering all of the extensions. 

In Tables \ref{tab:tomita_mem_table}, \ref{tab:tomita_var_table}, and \ref{tab:tomita_pref_table}, we show the results for the Tomita languages experiment for our MemRePs algorithm and the membership query-only baseline (where the cost for asking a preference query is set to $\infty$), averaged across 20 trials. Note that we omit the comparison between the number of equivalence queries asked by each method since these numbers were the same for both. Overall, we notice that the tradeoff between membership and preference queries is apparent as the relative cost between the two increases. However, this tradeoff is more pronounced in some languages (languages \#4, 5, and 7) than other (languages \#1, 2, and 6).  An example language (language \#5) is further illustrated in 

We also note that the number of preference queries needed in some cases, such as in languages 4 and 7, dramatically increase with the increased cost ratio. The high variance for preference queries needed also indicates that the combination of the CMAB algorithm and atom selection process leaves room for improvement and consistency.

\begin{table}
    \centering
    \small

    \begin{tabular}{ |l|cc|cc|cc|cc|cc|cc|cc| } 
        \cline{2-15}
        \multicolumn{1}{c|}{} 
        & \multicolumn{2}{c|}{DFA 1}
        & \multicolumn{2}{c|}{DFA 2}
        & \multicolumn{2}{c|}{DFA 3}
        & \multicolumn{2}{c|}{DFA 4}
        & \multicolumn{2}{c|}{DFA 5}
        & \multicolumn{2}{c|}{DFA 6}
        & \multicolumn{2}{c|}{DFA 7}\\
        \multicolumn{1}{c|}{} & Ours & Base & Ours & Base & Ours & Base & Ours & Base & Ours & Base & Ours & Base & Ours & Base \\
        \cline{1-15}
        Cost=1   & 3.1 & 3.6 &
                 6.6 & 7.0 & 
                 5.1  & 7.3 & 
                 12.5 & 16.6 & 
                 8.4 & 11.4 & 
                 5.7 & 6.7 & 
                 17.3 & 18.7 \\
                 
        Cost=2    & 3.1 & 3.6 &
                  6.2 & 7.0 & 
                  4.7 & 7.3 & 
                  11.2 & 16.6 & 
                  6.6 & 11.4 & 
                  5.5 & 6.7 & 
                  16.1 & 18.7 \\
                  
        Cost=4  & 3.0 & 3.6 & 
                  5.3 & 7.0 & 
                  3.8 & 7.3 & 
                  11.4 & 16.6 & 
                  5.6 & 11.4 & 
                  5.5 & 6.7 & 
                  14.9 & 18.7 \\
                  
        Cost=8   & 3.0 & 3.6 & 
                  3.5 & 7.0 & 
                  2.7 & 7.3 & 
                  6.8 & 16.6  & 
                  4.2 & 11.4 & 
                  5.4 & 6.7 & 
                  12.2 & 18.7\\
        \cline{1-15}
    \end{tabular}
    
    \caption{Mean number of membership queries asked by our membership-and-preference selection algorithm (Ours) in comparison to the membership-only baseline (Base) on the seven Tomita DFAs. Little to no difference is seen in the simpler DFAs, whereas the discrepancy in query amount is more pronounced in more complicated DFAs.}
    \label{tab:tomita_mem_table}

\end{table}

\begin{table}
    \centering
    \small
    \begin{tabular}{ |l|cc|cc|cc|cc|cc|cc|cc| }
        \cline{2-15}
        \multicolumn{1}{c|}{} 
        & \multicolumn{2}{c|}{DFA 1}
        & \multicolumn{2}{c|}{DFA 2}
        & \multicolumn{2}{c|}{DFA 3}
        & \multicolumn{2}{c|}{DFA 4}
        & \multicolumn{2}{c|}{DFA 5}
        & \multicolumn{2}{c|}{DFA 6}
        & \multicolumn{2}{c|}{DFA 7}\\
        \multicolumn{1}{c|}{} & Ours & Base & Ours & Base & Ours & Base & Ours & Base & Ours & Base & Ours & Base & Ours & Base \\
        \cline{1-15}
        Cost=1   & 0.13 & 0.23 & 
                  0.74 & 0 & 
                  0.66 & 0.49 & 
                  1.57 & 1.61 & 
                  1.28 & 1.01 & 
                  0.67 & 0.50 & 
                  2.00 & 1.78 \\
                   
        Cost=2    & 0.43 & 0.23 & 
                    1.01 & 0 & 
                    0.61 & 0.49 & 
                    1.62 & 1.61 & 
                    1.18 & 1.01 & 
                    0.92 & 0.50 & 
                    0.87 & 1.78 \\
                    
        Cost=4  & 0.22 & 0.23 & 
                  0.78 & 0 & 
                  0.87 & 0.49 & 
                  1.58 & 1.61 & 
                  0.64 & 1.01 & 
                  0.81 & 0.50 & 
                  2.19 & 1.78 \\
                  
        Cost=8   & 0.41 & 0.23 & 
                  0.67 & 0 & 
                  0.64 & 0.49 & 
                  0.97 & 1.61 & 
                  1.10 & 1.01 & 
                  0.49 & 0.50 & 
                  1.30 & 1.78\\
        \cline{1-15}
    \end{tabular}
    \caption{Variances for number of membership queries asked by our membership-and-preference selection algorithm in comparison to the membership-only baseline on the seven Tomita DFAs.}
    \label{tab:tomita_var_table}
\end{table}

\begin{table}
    \centering
    \small
    \begin{tabular}{ |l|cc|cc|cc|cc|cc|cc|cc| }
        \cline{2-15}
        \multicolumn{1}{c|}{} 
        & \multicolumn{2}{c|}{DFA 1}
        & \multicolumn{2}{c|}{DFA 2}
        & \multicolumn{2}{c|}{DFA 3}
        & \multicolumn{2}{c|}{DFA 4}
        & \multicolumn{2}{c|}{DFA 5}
        & \multicolumn{2}{c|}{DFA 6}
        & \multicolumn{2}{c|}{DFA 7}\\
        \multicolumn{1}{c|}{} & Mean & Var. & Mean & Var.  & Mean & Var. & Mean & Var.  & Mean & Var.  & Mean & Var.  & Mean & Var. \\
        \cline{1-15}
        Cost=1  & 1.4 & 0.75 & 
                  4.7 & 1.11 & 
                  4.6 & 2.72 & 
                  10.5 & 6.42 & 
                  5.1 & 2.14 & 
                  3.3 & 1.10 & 
                  11.5 & 4.18 \\
        
        Cost=2  & 2.6 & 0.88 & 
                  5.2 & 0.92 & 
                  5.0 & 1.66 & 
                  25.1 & 11.1 & 
                  7.0 & 1.76 & 
                  3.3 & 2.28 & 
                  22.0 & 9.27 \\
        
        Cost=4  & 3.1 & 0.79 & 
                  6.3 & 0.83 & 
                  9.7 & 2.05 & 
                  41.3 & 12.71 & 
                  8.1 & 2.95 & 
                  3.7 & 1.44 & 
                  44.3 & 8.22 \\
        
        Cost=8  & 3.8 & 0.83 & 
                  8.1 & 0.77 & 
                  15.9 & 2.28 & 
                  84.3 & 17.23 & 
                  9.1 & 5.11 & 
                  4.1 & 1.34 & 
                  78.7 & 16.65\\
        \cline{1-15}
    \end{tabular}
    \caption{Mean and variance for number of preference queries asked by our membership-and-preference selection algorithm on the seven Tomita DFAs.}
    \label{tab:tomita_pref_table}
\end{table}

\begin{figure}
  \centering
  
  \scalebox{0.9}{
  \includegraphics[width=0.5\textwidth,height=0.3\textheight,keepaspectratio]{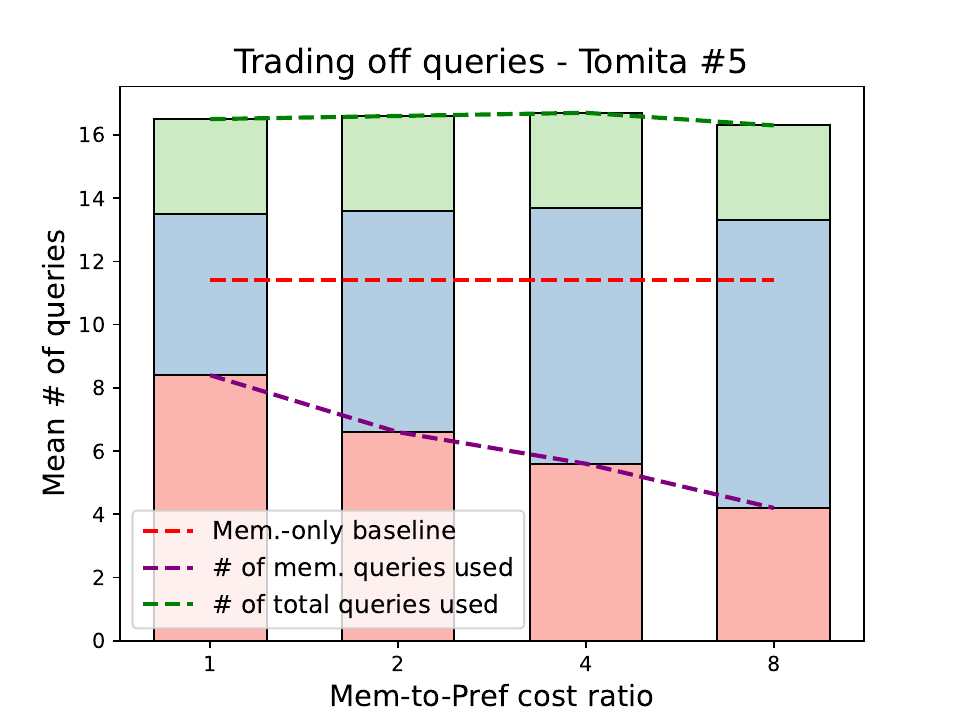}
  }
  \caption{Trade-off between preference
    queries and membership queries for tomita language \#5. The bars plotted show the contribution of
    membership (red, bottom), preference (blue,
    middle), and equivalence (green, top) queries.}
  \label{fig:tomita_tradeoff}
\end{figure}

\subsubsection{Robustness Experiment}\label{sec:robustnessexp}
\begin{figure}
  \centering
  
  \scalebox{0.9}{
  \includegraphics[width=0.5\textwidth,height=0.3\textheight,keepaspectratio]{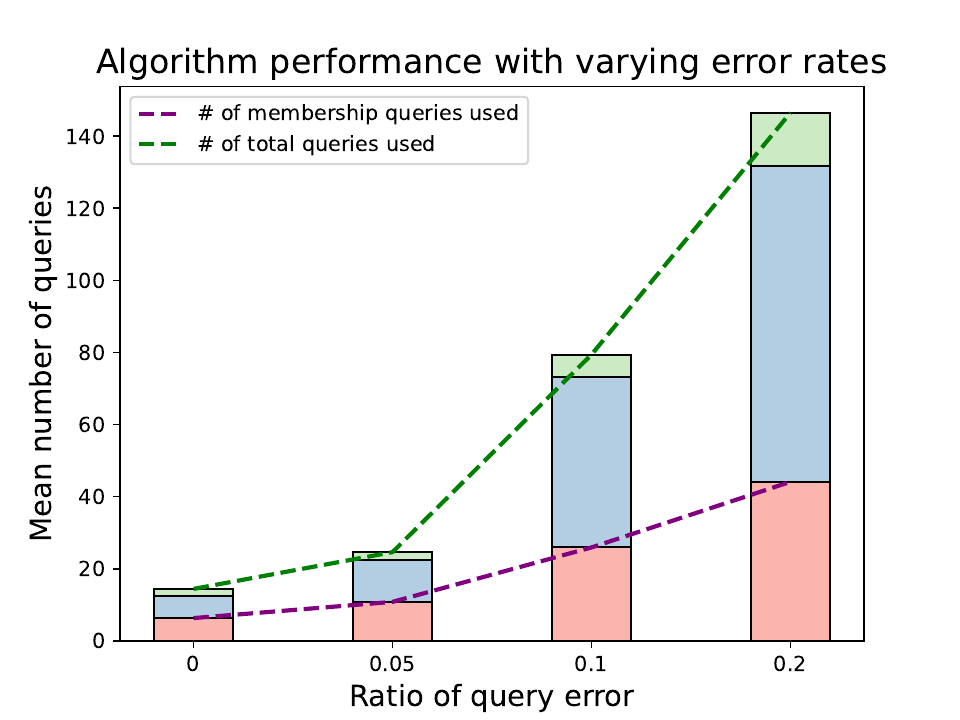}
  }
  \caption{Greater numbers of queries are required to learn the target DFA (Tomita language \#6) as the error rate increases. The bars plotted show the contribution of
    membership (red, bottom), preference (blue,
    middle), and equivalence (green, top) queries.}
  \label{fig:robustness}
\end{figure}

As mentioned in the main text, we designed our algorithm to be robust in noisy settings, where the response to a query is flipped to be incorrect some proportion of the time. In our algorithm's implementation for DFAs, if a labeling error occurs that causes a violation, the UNSAT-core is extracted to see which existing assumptions caused this violation, and those assumptions are then dropped. We demonstrate the effect of labeling errors in an experiment, where an example DFA (in this case, Tomita language \#6) is learned in settings of increasing proportions of labeling errors. The results are displayed in figure \ref{fig:robustness}, where an increasing rate of error causes more assumptions that need to be dropped by our algorithm, resulting in more queries required to learn the correct concept. In other words, as errors are made more frequently and assumptions are discarded more often, it becomes harder for the necessary set of assumptions to be efficiently obtained by the learner.

\subsubsection{Scalability Experiments}\label{sec:scalability_tomita}
To understand how our algorithm scales as our target DFA increases in complexity, we evaluate our algorithm's scalability on a simple one-symbol language that determines whether the length of an input sequence is modulo some positive integer $k$. The size of the DFA is $k$ states with a single accepting state. We provide the results of our algorithm's performance as $k$ increases in Table~\ref{tab:scalability_modulo}.  Not included in the table is the number of equivalence queries used in each DFA, which did not vary as a function of membership-to-preference cost ratio. The mean number of equivalence queries asked were 6.4, 9.1, 9.9, and 22.6 for DFA states 5, 10, 20, and 40, respectively. The equivalence queries in this setting were highly informative, allowing the number of other query types to scale efficiently but somewhat restricting those queries' utility.

\begin{table} 
    \centering
    \small
    \begin{tabular}{ |l|cc|cc|cc|cc| }
        \cline{2-9}
        \multicolumn{1}{c|}{} 
        & \multicolumn{2}{c|}{5 States}
        & \multicolumn{2}{c|}{10 States}
        & \multicolumn{2}{c|}{20 States}
        & \multicolumn{2}{c|}{40 States}\\
        \multicolumn{1}{c|}{} & \# Mem. & \# Pref. & \# Mem. & \# Pref. & \# Mem. & \# Pref. & \# Mem. & \# Pref. \\
        \cline{1-9}
        Cost=1  & 2.2 & 1.2 & 
                  5.7 & 3.6 & 
                  9.4 & 7.4 & 
                  18.7 & 7.8 \\
        
        Cost=2  & 2.0 & 1.5 & 
                  4.7 & 5.6 & 
                  8.1 & 8.4 & 
                  17.9 & 9.9 \\
        
        Cost=4  & 1.4 & 2.6  & 
                  4.6 & 6.6 & 
                  6.2 & 12.2 & 
                  15.4 & 19.5 \\
        
        Cost=8  & 1.0 & 2.9 & 
                  3.9 & 10.9 & 
                  3.7 & 17.8 & 
                  11.3 & 31.4 \\
        \cline{1-9}
    \end{tabular}
    \caption{Number of Membership and Preference Queries for the scaled modulo DFA structure, averaged over ten trials.}
    \label{tab:scalability_modulo}
\end{table}

In addition to the previous experiment, we generalized Tomita Language \#4, which originally is defined as a 3-state DFA that encodes the task ``any string without more than 2 consecutive `0's'', to any string with more than $n$ consecutive `0's. The size of the DFA in states is $n + 2$, including the rejecting sink state. We vary $n$ from 1 to 4 and present our results in Table~\ref{tab:scalability_tomita}. We note that the number of queries required quickly increases with the increase in number of states, which is to be expected given the super-linear increase in search space size with number of states.

\begin{table*} 
    \centering
    \small
    \begin{tabular}{ |l|cc|cc|cc|cc| }
        \cline{2-9}
        \multicolumn{1}{c|}{} 
        & \multicolumn{2}{c|}{3 States}
        & \multicolumn{2}{c|}{4 States}
        & \multicolumn{2}{c|}{5 States}
        & \multicolumn{2}{c|}{6 States}\\
        \multicolumn{1}{c|}{} & \# Mem. & \# Pref. & \# Mem. & \# Pref. & \# Mem. & \# Pref. & \# Mem. & \# Pref. \\
        \cline{1-9}
        Cost=1  & 7.3 & 3.1 & 
                  12.5 & 10.5 & 
                  23.3 & 14.3 & 
                  32.4 & 34.6 \\
        
        Cost=2  & 6.5 & 6.3 & 
                  11.2 & 25.1 & 
                  20.3 & 26.9 & 
                  28.6 & 59.0 \\
        
        Cost=4  & 5.6 & 17.7  & 
                  11.4 & 41.3 & 
                  19.1 & 71.7 & 
                  25.8 & 143.6 \\
        
        Cost=8  & 5.1 & 34.1 & 
                  6.8 & 84.3 & 
                  17.9 & 147.2 & 
                  23.6 & 218.6 \\
        \cline{1-9}
    \end{tabular}
    \caption{Number of Membership and Preference Queries for the scaled Tomita \#4 experiment, averaged over ten trials. The number of equivalence queries remained constant over costs.}
    \label{tab:scalability_tomita}

\end{table*}

%% file: imgs/monotone_geometry.pdf_tex
\begingroup%
  \makeatletter%
  \providecommand\color[2][]{%
    \errmessage{(Inkscape) Color is used for the text in Inkscape, but the package 'color.sty' is not loaded}%
    \renewcommand\color[2][]{}%
  }%
  \providecommand\transparent[1]{%
    \errmessage{(Inkscape) Transparency is used (non-zero) for the text in Inkscape, but the package 'transparent.sty' is not loaded}%
    \renewcommand\transparent[1]{}%
  }%
  \providecommand\rotatebox[2]{#2}%
  \newcommand*\fsize{\dimexpr\f@size pt\relax}%
  \newcommand*\lineheight[1]{\fontsize{\fsize}{#1\fsize}\selectfont}%
  \ifx\svgwidth\undefined%
    \setlength{\unitlength}{292.2057724bp}%
    \ifx\svgscale\undefined%
      \relax%
    \else%
      \setlength{\unitlength}{\unitlength * \real{\svgscale}}%
    \fi%
  \else%
    \setlength{\unitlength}{\svgwidth}%
  \fi%
  \global\let\svgwidth\undefined%
  \global\let\svgscale\undefined%
  \makeatother%
  \begin{picture}(1,0.38559236)%
    \lineheight{1}%
    \setlength\tabcolsep{0pt}%
    \put(0.7216293,-0.00110124){\makebox(0,0)[t]{\lineheight{1.25}\smash{\begin{tabular}[t]{c}min distance\end{tabular}}}}%
    \put(0.91267127,0.19538051){\rotatebox{-90}{\makebox(0,0)[t]{\lineheight{1.25}\smash{\begin{tabular}[t]{c}max travel time\end{tabular}}}}}%
    \put(0,0){\includegraphics[width=\unitlength,page=1]{monotone_geometry.pdf}}%
    \put(0.13652028,0.00119578){\makebox(0,0)[t]{\lineheight{1.25}\smash{\begin{tabular}[t]{c}x\end{tabular}}}}%
    \put(0.42563893,0.19554813){\rotatebox{6.8989626}{\makebox(0,0)[t]{\lineheight{1.25}\smash{\begin{tabular}[t]{c}path to boundary\end{tabular}}}}}%
    \put(0,0){\includegraphics[width=\unitlength,page=2]{monotone_geometry.pdf}}%
    \put(0.72840189,0.09035213){\makebox(0,0)[t]{\lineheight{1.25}\smash{\begin{tabular}[t]{c}$\Theta^+_x$\end{tabular}}}}%
    \put(0.78780356,0.26291436){\makebox(0,0)[t]{\lineheight{1.25}\smash{\begin{tabular}[t]{c}$\theta$\end{tabular}}}}%
    \put(0,0){\includegraphics[width=\unitlength,page=3]{monotone_geometry.pdf}}%
  \end{picture}%
\endgroup%

%% file: errors.tex
Our concept learning algorithm can be adapted to gracefully handle two kinds of labeling errors: 
(i) \emph{Preorder violations} and (ii) \emph{MemRep violations}.
A preorder violation occurs when the underlying order relation is observed
to not be transitive or reflexive. This can be visualized using a
Hasse Diagram where such a violation would either correspond
to a cycle or an inconsistency in the node equivalence
classes. 
Similarly, a MemReP violation occurs when the
Hasse diagram contains an edge, $(x, y)$, where $x \prec y$, but
$x \in \concept^*$ and $y \notin \concept^*$.

Both classes of violations are easy to detect and isolate.
In an explicit Hasse diagram representation, a topological pass over the graph
suffices. In the case of SAT based concept classes, e.g., our DFA learning
experiment in Section~\ref{sec:applications}, we simply analyze the UNSAT-core to determine which queries must be dropped to find a consistent hypothesis. 
In either setting, one can alert the user, allowing the
violating query responses to be dropped or corrected before resuming the
learning algorithm. Combining with a final conformance tester which asks additional redundant queries from a test distribution, yields a probably approximately correct concept~\cite{DBLP:books/daglib/0041035}.